\documentclass[a4paper,10pt]{article}
\usepackage{color}
\usepackage{amsmath}    
\usepackage{amssymb}
\usepackage{graphicx}   
\usepackage{subfigure}

\title{Frustrated hierarchical synchronization and emergent complexity
  in the human connectome network.}

\author{Pablo Villegas, Paolo Moretti$^{*}$, and Miguel A. Mu\~noz,\\
\small{{Departamento de Electromagnetismo y F{\'i}sica de la
 Materia e}}\\ \small{ Instituto Carlos I de F{\'i}sica Te{\'o}rica y
 Computacional.} \\ \small{Universidad de Granada, E-18071 Granada, Spain}
 \\ \small{$^{*}$Corresponding author: moretti.paolo@gmail.com}}

\begin{document}     

\maketitle
\begin{abstract}
  The spontaneous emergence of coherent behavior through
  synchronization plays a key role in neural function, and its
  anomalies often lie at the basis of pathologies. Here we employ a
  parsimonious (mesoscopic) approach to study analytically and
  computationally the synchronization (Kuramoto) dynamics on the
  actual human-brain connectome network.  We elucidate the existence
  of a so-far-uncovered intermediate phase, placed between the
  standard synchronous and asynchronous phases, i.e.  between order
  and disorder. This novel phase stems from the hierarchical modular
  organization of the connectome.  Where one would expect a
  hierarchical synchronization process, we show that the interplay
  between structural bottlenecks and quenched intrinsic frequency
  heterogeneities at many different scales, gives rise to frustrated
  synchronization, metastability, and chimera-like states, resulting in a
  very rich and complex phenomenology.  We uncover the origin of the
  dynamic freezing behind these features by using spectral graph
  theory and discuss how the emerging complex synchronization patterns
  relate to the need for the brain to access --in a robust though
  flexible way-- a large variety of functional attractors and
  dynamical repertoires without {\it ad hoc} fine-tuning to a critical
  point.
\end{abstract}

Neuro-imaging techniques have allowed the reconstruction of structural
human brain networks, composed of hundreds of neural regions and
thousands of white-matter fiber interconnections.  The resulting
``human connectome'' (HC) \cite{Hagmann,Honey09} turns out to be
organized in moduli --characterized by a much larger intra than inter
connectivity-- structured in a hierarchical nested fashion across many
scales
\cite{Bullmore-Sporns,Sporns,Review-Kaiser,Review-Bullmore,Buzsaki,Zhou06,
Ivkovic,Sporns2014}.
On the other hand, ``functional'' connections between nodes in these
networks have been empirically inferred from correlations in neural
activity as detected in electroencephalogram and functional magnetic
resonance time series. Unveiling how structural and functional
networks influence and constrain each other is a task of outmost
importance. A few pioneering works found that the hierarchical-modular
organization of structural brain networks has profound implications
for neural dynamics
\cite{Zhou06,Zhou07,Kaiser07,Kaiser10,Rubinov}. For example, neural
activity propagates in hierarchical networks in a rather distinctive
way, not observed on simpler networks \cite{Nat-Comm}; beside the
usual two phases --percolating and non-percolating-- commonly
encountered in models of activity propagation, an intermediate
``Griffiths phase'' \cite{Vojta-Review} emerges on the hierarchical HC
network \cite{GPCN,Nat-Comm,GPCN2}.  Such a Griffiths phase stems from the
existence of highly diverse relatively-isolated moduli or ``rare
regions'' where neural activity remains mostly localized generating slow
dynamics and very large responses to perturbations \cite{Vojta-Review,Nat-Comm,GPCN,GPCN2}.

Brain function requires coordinated or coherent neural activity at a
wide range of scales, thus, neural synchronization is a major theme in
neuroscience \cite{Buzsaki,Bennett2004,Breakspear-multiscale}.
Synchronization plays a key role in vision \cite{Sompolinsky88},
memory \cite{Klimesch1996}, neural communication
\cite{Deco-inf-transf}, and other cognitive functions
\cite{Niebur2000}. An excess of synchrony results in pathologies such
as epilepsy or Parkinsonian disease, while neurological deficit of
synchronization has been related to autism and schizophrenia
\cite{Kandel00}.

Our aim here is to scrutinize the special features of synchronization
dynamics \cite{RPK-book} --as exemplified by the canonical Kuramoto
model \cite{Kuramoto75,Strogatz00,Acebron-Review}-- running on top of
the best available human connectome mapping
\cite{Hagmann,Honey09,Arenas-Review}. This consists of a network of
$998$ nodes, each of them representing a mesoscopic population of
neurons --able to produce self-sustained oscillations
\cite{Sporns2011}-- whose mutual connections are encoded by a
symmetric weighted connectivity matrix $\bf{W}$
\cite{Hagmann,Honey09}.  The validity of this admittedly simplistic
Kuramoto model as a convenient tool to explore the generic features of
complex brain dynamics at a large scales has been recently emphasized
in the literature \cite{Breakspear2010,Sporns2011,Arenas10cat}.  Here,
we uncover the existence of a novel intermediate phase for
synchronization dynamics --similar in spirit to the Griffiths phases
discussed above-- which stems from the hierarchical modular
organization of the HC and which gives rise to very complex and rich
synchronization dynamical patterns.  We identify this novel phase as
the optimal regime for the brain to harbor complex behavior, large
dynamical repertoires, and optimal trade-offs between local
segregation and global integration.

The Kuramoto dynamics on a generic network (see \cite{Arenas-Review}
for a nice and comprehensive review) is defined by:
\begin{equation}
\dot{\theta}_{i}(t)=\omega_{i}+ k \sum_{j=1}^{N}
W_{ij}\sin\left[\theta_{j}(t)-\theta_{i}(t)\right], \label{eq:Kuramoto}
\end{equation}
where ${\theta}_{i}(t)$ is the phase at node $i$ at time $t$. The
intrinsic frequencies $\omega_{i}$ --accounting for node
heterogeneity-- are extracted from some arbitrary distribution
function $g(\omega)$, $W_{ij}$ are the elements of the $N \times N$
connectivity matrix $\bf{W}$, and $k$ is the coupling strength.  Time
delays, noise, and phase frustration could also be straightforwardly
implemented.  The Kuramoto order parameter is defined as
$Z(t)=R(t)\mathrm{e}^{i\psi(t)} = \langle \mathrm{e}^{i \theta_i(t)}
\rangle$, where $0 \leq R(t) \leq 1$ gauges the overall coherence and
$\psi(t)$ is the average global phase.  In large populations of
well-connected oscillators  without frequency dispersion,
perfect coherence ($R =1$) emerges for any coupling strength; on the
other hand, frequency heterogeneity leads to a phase transition at
some critical value of $k$, separating a coherent steady state from an
incoherent one
\cite{Kuramoto75,Strogatz00,Acebron-Review,Arenas-Review}.  Analytical
insight onto this phase transition can be obtained using the
celebrated Ott-Antonsen (OA) ansatz, allowing for a projection of the
high-dimensional dynamics into an evolution equation for $Z(t)$ with
remarkable accuracy in the large-$N$ limit \cite{OA,Skardal-Restrepo}.
\begin{figure}[htbp]
\centering
\includegraphics[width=12cm,angle=0]{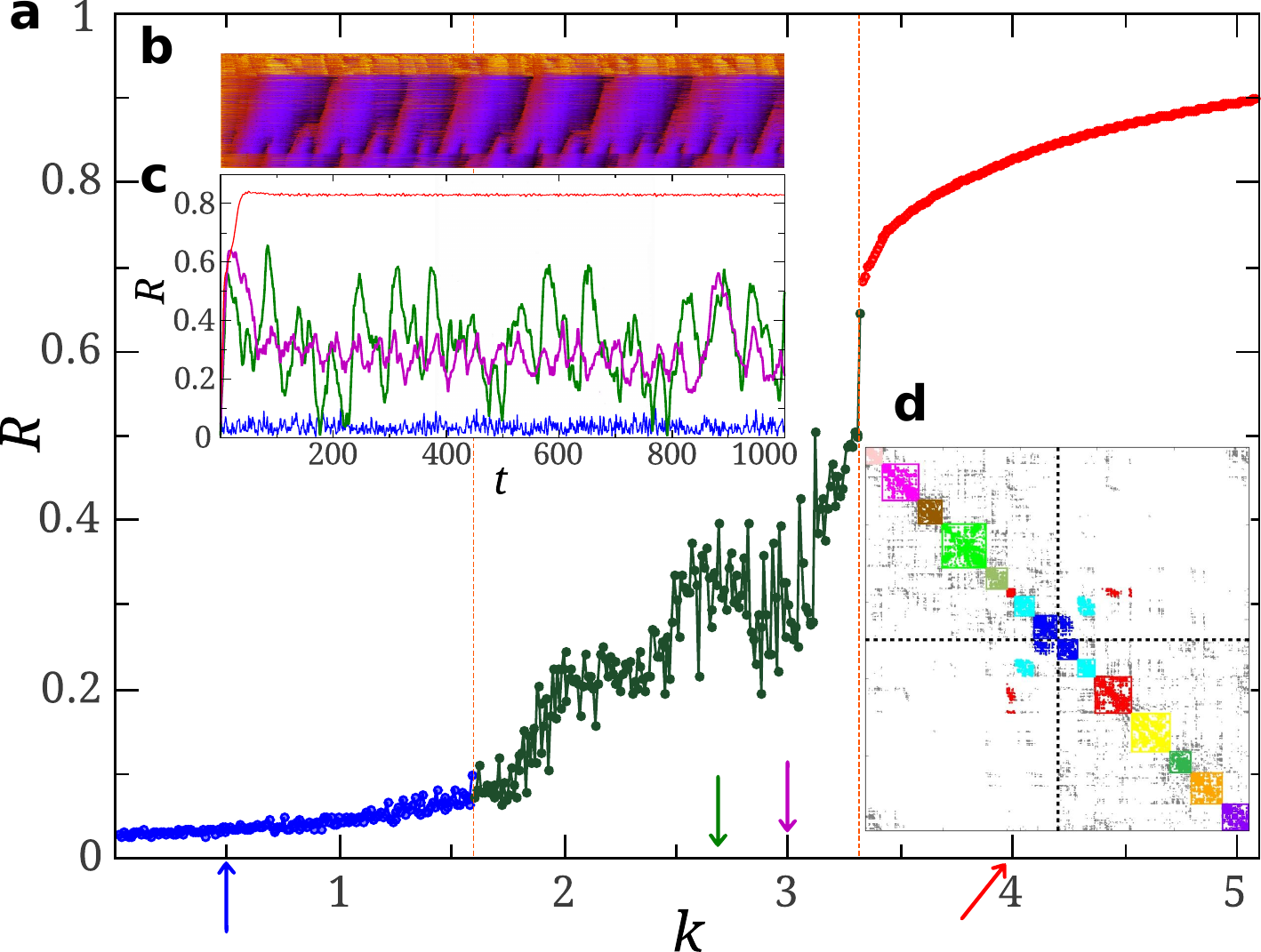}
\caption{Global synchronization dynamics in the human connectome.
  a) Time average of the order parameter $R(t)$,
  for Kuramoto dynamics on the HC network for a specific and fixed set
  of frequencies extracted from a $N(0,1)$ Gaussian distribution. A
  broad intermediate regime separates the incoherent phase (low $k$)
  from the synchronous one (high $k$). In this regime, coherence
  increases with $k$ in an intermittent fashion, and with strong
  dependence on the frequency realization.  b) Raster plot of
  individual phases (vertical axis) showing local rather than global
  synchrony and illustrating the coexistence of coherent and
  incoherent nodes ($k=2.7$) as time runs.  c) $R(t)$ for $4$ values
  of $k$ (arrows in the main plot).  d) Adjacency matrix of the HC
  network with nodes ordered to emphasize its modular structure as
  highlighted by a community detection algorithm (main text), keeping
  the partition into the $2$ hemispheres (dashed lines). Intra-modular
  connections (shown in color) are dense while inter-modular ones
  (grey) are limited to tiny subsets, acting as interfaces between
  moduli.  Integration between hemispheres is mostly carried out by
  the $3$ central moduli. This plot visually illustrates the
  hierarchical modular organization of the human connectome network.}
\end{figure} 
\section*{Results}
\subsection*{Novel phase between order and disorder in the HC}
We have performed a computational study of the Kuramoto model running
on top of the HC network (details are given in the Methods section).
Our results reveal the existence of an intermediate regime placed
between the coherent and the incoherent phase (see Fig.1).  This is
characterized by broad quasi-periodic temporal oscillations of $R(t)$
which wildly depend upon the realization of intrinsic frequencies
\cite{Arenas1,Arenas2}.  Anomalously large sampling times would be
required to extract good statistics for the actual mean values and
variances.  Collective oscillations of $R(t)$ are a straightforward
manifestation of partial synchronization and they are robust against
changes in the frequency distribution (e.g. Gaussian, Lorentzian,
uniform, etc.)  whereas the location and width of the intermediate
phase depend upon details.  As this phenomenology is reminiscent of
Griffiths phases --posed in between order and disorder and stemmig
from the existence of semi-isolated regions
\cite{Vojta-Review,GPCN,Nat-Comm}-- it is natural to investigate how
the HC hierarchical modular structure affects synchronization
dynamics.

Any network with perfectly isolated and independently synchronized
moduli trivially exhibits oscillations of $R(t)$, with amplitude
peaking at times when maximal mutual synchronization happens to be
incidentally achieved. Such oscillations can become chaotic if a
finite and relatively small number of different coherent moduli are
coupled together \cite{Popovych}. Thus, in a connected network without
delays or other additional ingredients, oscillations in the global
coherence are the trademark of strong modular structure with weakly
interconnected moduli.

Strong modular organization into distinct hierarchical levels is
indeed present in the HC as reveled by standard community detection
algorithms \cite{Radatools,Ivkovic} and as already discussed in the
literature (see e.g. \cite{Nat-Comm} and references therein).  For
instance, we have found that the optimal partition into disjoint
communities --i.e. the partition maximizing the modularity parameter
\cite{Newman-Review}-- corresponds to a division in $12$ communities
(see Fig.1d) while, at a higher hierarchical level, a separation into
just $2$ moduli --the $2$ cerebral hemispheres-- is obtained
\cite{Honey09} (Fig.1d).  Obviously these $2$ coarser moduli include
the $12$ above as sub-moduli.  Although more levels of hierarchical
partitioning could be inferred (see e.g. \cite{Sporns2014} and
refs. therein), for the sake of simplicity we focus on these two
levels $l$, $l=1$ and $l=2$ with $12$ and $2$ moduli, respectively.
\begin{figure}[ht]
\begin{center}
\vspace{-0cm}
   \includegraphics[width=12cm]{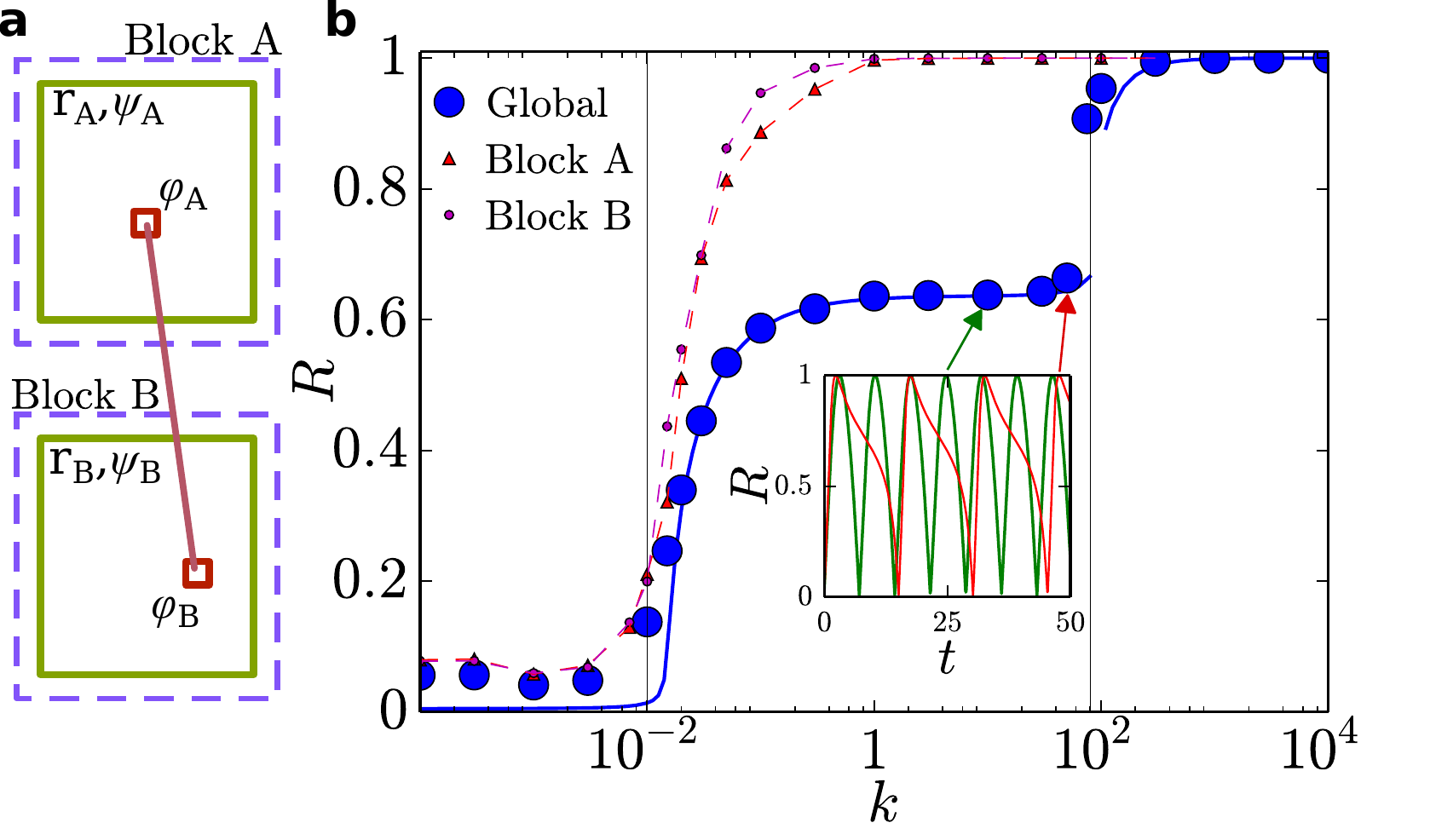}
   \caption{Two-block model.
     (a) Sketch of the two-block model.  (b)
     Global order parameter for the two-block model with
     $M=128$ and two interfacial nodes. 
     Results of the numerical integration of the $258$
     Kuramoto equations (blue points) are in strikingly good agreement
     with the integration of Eqs.(\ref{OA}) (solid blue line). Local
     block-wise order parameters are shown for comparison (small
     symbols; dashed lines are guides to the eye).  A first
     transition, where local order emerges, occurs at $k\approx 0.02$,
     while global coherence is reached at $k \approx 90$.  In the
     intermediate region, $R(t)$ oscillates (inset), revealing the
     lack of global coherence.  Despite the simplicity of this toy
     model, these results constitute the essential building-block upon
     which further levels of complexity rely (see main text).}
    \label{fig:blocks}
\end{center}
\end{figure}

\subsection*{A simplistic model for global oscillations}
To shed further light on the properties of synchronization on the HC,
we consider a very simple network model --allowing for analytical
understanding-- which will constitute the elementary
``building-block'' for subsequent more complex analyses.  This
consists of a few blocks with very large internal connectivity and
very sparse inter-connectivity. Each block is composed by a bulk of $M
\gg 1$ nodes that share no connection with the outside and a
relatively small ``interfacial'' set that connects with nodes in other
blocks. For instance, in the simplest realization, consisting of just
two blocks connected by a single pair of nodes (Fig.2), each block is
endowed with local coherence $r_{\mathrm{A},\mathrm{B}}$, average
phase $\psi_{\mathrm{A},\mathrm{B}}$, and average characteristic
frequency $\omega_\mathrm{A,B}$, while 1-node interfaces have perfect
coherence $r=1$, phase $\varphi_{\mathrm{A},\mathrm{B}}$, and
characteristic frequency $\nu_{\mathrm{A},\mathrm{B}}$.  In this case,
$N=2M+2$, and the OA ansatz can be safely applied to each block (large
$M$) but not to single-node interfaces.  In the particular case
(convenient for analytical treatment) in which $g(w)$ are zero-mean
Lorentz distributions $g(\omega) = \frac{1}{\pi}
\frac{\delta}{(\omega-\Omega_0)^2 + \delta^2}$ with spreads
$\delta_{\mathrm{A},\mathrm{B}}$, the resulting set of OA equations
can be easily shown to be:
 \begin{eqnarray}
  \dot{\psi}_\mathrm{A} & =& 
  \omega_{\mathrm{A}} + k
  \frac{1+r_\mathrm{A}^2}{2r_\mathrm{A}} 
  \sin(\varphi_\mathrm{A}-\psi_\mathrm{A}) \nonumber \\
  \dot{r}_\mathrm{A}  &=&
  -\delta_\mathrm{A}r_\mathrm{A} + k
  \frac{1-r_\mathrm{A}^2}{2} 
  \left[Mr_\mathrm{A}+ 
  \cos(\varphi_\mathrm{A}-\psi_\mathrm{A})\right] \nonumber \\
\dot{\varphi}_\mathrm{A} & = &
  \nu_\mathrm{A}+
k \left[
Mr_\mathrm{A}\sin(\psi_\mathrm{A}-\varphi_\mathrm{A})+\sin(\varphi_\mathrm{B}
    -\varphi_\mathrm{A})\right] 
\label{OA}
\end{eqnarray}
(together with $r=1$ for each 1-node interface), and a symmetric set ($A
\leftrightarrow B$) for block $\mathrm{B}$.  The solution of Eq.(2)
--displayed in Fig.2-- reveals a transition to local coherence within
each block at a certain threshold value of $k\approx 0.02$. As soon as
local order is attained, $r_\mathrm{A,B}\approx 1$ and
$\dot{\psi}_\mathrm{A,B}\approx 0$, from Eq.(2) the mutual
synchronization process obeys
 \begin{equation}
 \dot{\varphi}_\mathrm{A}\approx
 (\nu_\mathrm{A}+M\omega_{\mathrm{A}})+
 k\sin\left(\varphi_\mathrm{B}-\varphi_\mathrm{A}\right) 
 \end{equation}
 and a symmetrical equation for $\dot{\varphi}_\mathrm{B}$. For small
 $k$, the right-hand side is dominated by
 $\nu_\mathrm{A}+M\omega_{\mathrm{A}}$: whereas the average value
 $\omega_\mathrm{A}$ becomes arbitrarily small within blocks (assuming
 that $M$ is large), the frequency $\nu_\mathrm{A}$ does not.
 Consequently, synchronization between the two blocks through the
 interfacial link is frustrated: each block remains internally
 synchronized but is unable to achieve coherence with the other over a
 broad interval of coupling strengths. This interval is delimited
 above by a second transition at $k \sim
 \max\{M|\omega_\mathrm{A,B}|,\nu_{\mathrm{A,B}}\}$, where $k$ is
 large enough as to overcome frustration and generate global
 coherence. This picture is confirmed by numerical integration of the
 full system of $N$ coupled Kuramoto equations as well as by its OA
 approximation (Eq.(2)), both in remarkably good agreement. Therefore,
 local and global coherences have their onsets at two well-separated
 transition points \cite{Skardal-Restrepo} and --similarly to the much
 more complex HC case-- $R$ oscillates in the intermediate regime
 (Fig.2).  Similar results hold for versions of the model with more
 than two moduli (e.g. $4$; see below).  The existence of two distinct
 (local and global) transitions had already been reported in a recent
 study of many blocks with much stronger inter-moduli connections than
 here \cite{Skardal-Restrepo} (even if, owing to this difference, no
 sign of an intermediate oscillatory phase was reported).
 In particular, the value of two-block models has already been
 explored in the past, for systems of identical oscillators with
 non-zero phase lags, in which each node is coupled equally to all the
 others in its community, and less strongly to those in the other
 \cite{Abrams2008}. In such systems, local coherence emerged for large
 enough values of the phase lag.
 Our two-block model shows that the presence of ``structural
 bottlenecks'' between moduli combined with heterogeneous frequencies
 at their contact nodes (interfaces) are essential ingredients to
 generate a broad region of global oscillations in $R$, even in the
 absence of phase lag.  Still, it is obviously a too-simplistic model
 to account for all the rich phenomenology emerging on the HC, as we
 show now.
\begin{figure}[htbp]
\centering
\includegraphics[width=12cm,angle=0]{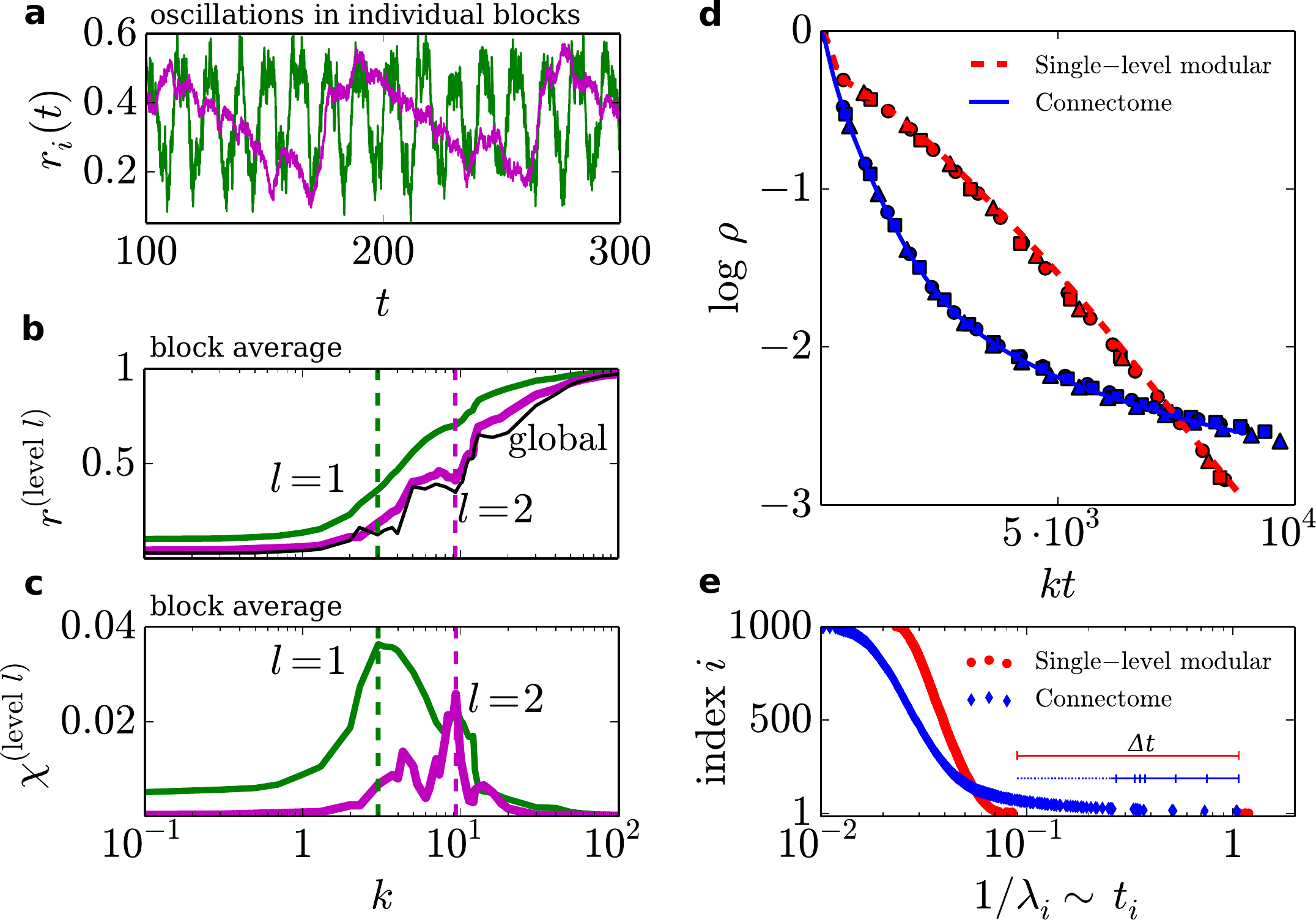}
\caption{Local synchronization in the human connectome.
  a) Oscillations of the local order parameters
  (``chimera-like states'') in one particular modulus in the partitions of
  the HC into $12$ (green, $l=1$, and $k=3$) and $2$ (magenta, $l=2$,
  and $k=10$) moduli, respectively. The characteristic frequency of
  these oscillations is typically between $0.01$ and $0.1 Hz$ (a range
  which coincides with slow modes detected in brain activity; see
  e.g. \cite{Breakspear2010}).  b) Average of the local order
  parameter over all moduli and c) chimera index for moduli at levels
  as in a), as a function of $k$.  Global order (thin black line in
  b)) emerges only after local order is attained at lower levels. d)
  Average decay of activity $\rho$ for identical frequencies
  $\omega=0$ in the HC network and comparison with a single-level
  modular network (made up of $4$ similar random moduli at a single
  hierarchical level) of the same size and average connectivity as the
  HC network. Symbols stand for different values of $k$.  e)
  Characteristic decay times corresponding to the inverse of the first
  $1000$ non-trivial eigenvalues of the Laplacian matrix (x axis) as a
  function of their respective ordered indices (y axis), for networks
  as in d). The stretched exponential behavior in d) is the result of
  the convolution of slow time scales associated with small
  eigenvalues in e).}
\end{figure}
\subsection*{Oscillations of local coherence in the HC}
Fig.3 shows numerical results for the local order parameter $r^{(l)}$
for some of the moduli at the $2$ hierarchical levels, $l=1$ and $l=2$
in the HC network.  It reveals that (Fig.3a) local coherences exhibit
oscillatory patterns in time (with characteristic frequencies
typically between $0.01$ and $0.1 Hz$) and that (Fig.3b) the
transition to local coherence at progressively higher hierarchical
level occurs at progressively larger values of $k$; i.e. coherence
emerges out of a hierarchical bottom-up process as illustrated above
for the for the two-block model (see
\cite{Arenas_scales,Skardal-Restrepo}).  Observe, however, that local
oscillations were not present in the two-block model.  This suggests
that the $12$ moduli in the HC are on their turn composed of finer
sub-moduli and that structural frustration, as introduced above,
affects all hierarchical levels.  The average variance of local
coherences (called chimera index, $\chi$) \cite{Shanahan2010} exhibits
a marked peak --reflecting maximal configurational variability-- at
the transition point for the corresponding level (Fig. 3b-c and
Methods section).  Similar intra-modular oscillatory patterns --dubbed
{\it chimera states}-- have been recently found
\cite{Abrams2008,Abrams2004,Shanahan2010,Shanahan2012} in Kuramoto 
models in which explicit phase lags induce a different kind of
frustration, hindering global synchronization. Strictly speaking, chimeras
are defined in systems of identical oscillators. In such a case, a non-zero
phase lag term is essential for partial synchronization to occur. 
Realistic models of the brain, however, require oscillators to be 
heterogenous. States of partial synchronization in empirical brain networks 
with frequency heterogeneity have been found for Kuramoto models with explicit 
time delays \cite{Sporns2011}. In contrast, the
chimera-like states put forward here have a purely structural origin, as they
arise from the network topology. It was noted in the past that synchronization 
in a synthetic network with hubs could be limited to those hubs by tuning 
clustering properties, and global order could be attained in a monotonous 
step-like fashion upon increasing $k$ \cite{McGraw2005}.
Fig.3b instead reveals that the ordering process in the hierarchical modular HC 
may be 
non-monotonous: coherence does not systematically grow with $k$. Indeed, the 
emergence of local order in some community may hinder or reduce coherence in
others, inducing local ``desynchronization'' and reflecting the
metastable nature of the explored states.

\subsection*{Anomalous dynamics on the HC} Fixing all intrinsic
frequencies to be identical allows us to focus specifically on
structural effects. Thus, we consider, without loss of generality, the
simple case $\omega_i=0$, and define the ``activity'' $\rho=1-\langle
R\rangle$. In this case, perfect asymptotic coherence should emerge
for all values of $k$ but, as illustrated in Fig. 3d, the convergence
towards $\rho=0$ turns out to be extremely slow (much slower than
exponential). This effect can be analytically investigated assuming
that, for large enough times, all phase differences are relatively
small. Then, up to first order, $\dot{\theta}_i = - k \sum_j
L_{ij}\theta_j$ where $L_{ij}=\delta_{ij}\sum_l W_{jl}-W_{ij}$ are the
elements of the Laplacian matrix \cite{FanChung,JStat2006}. Solving the linear 
problem,
$\theta_i(t)= \sum_{l,j}\mathrm{e}^{- k \lambda_l
  t}v^l_iv^l_j\theta_j(0)$, where $\lambda_l$ denotes the $l$-th
Laplacian eigenvalue ($0=\lambda_1 < \lambda_2 < ... < \lambda_N$) and
$v^l_i$ the $i$-th component of the corresponding eigenvector. Given
that the averaged order parameter can be written as $Z (t) \approx \sum_j
(1+\mathrm{i}\theta_j-\frac{1}{2}\theta_j^2)/N$, averaging over
initial conditions, and considering that (as the Laplacian has zero
row-sums \cite{JStat2006}) $\lambda_1=0$, we obtain 
\begin{equation}\label{eq:discrete_lambda}
\rho(t) =
\frac{\sigma^2}{2}\sum_{l=2}^N \mathrm{e}^{-2 k \lambda_l t},
\end{equation}
where $\sigma$ is the standard deviation of the initial phases. This
expression holds for any connected network.  As usual, the larger the
spectral gap $\lambda_2$, the more ``entangled'' \cite{JStat2006} the
network and thus the more difficult to divide it into well separated
moduli ($\lambda_2=0$ only for disconnected networks)
\cite{JStat2006,FanChung}.  For large spectral gaps all timescales are
fast, and the last expression can be approximated by its leading
contribution, ensuing exponential relaxation to $\rho=0$, as in fact
observed in well-connected network architectures (Erd\H os-R\'enyi,
scale free, etc. \cite{Newman-Review}). This is not the case for the
HC matrix, for which a tail of small non-degenerate eigenvalues is
encountered (see Fig.3e and \cite{Nat-Comm}).  Each eigenvalue
$\lambda_i$ in the tail corresponds to a natural division of moduli
into sub-moduli \cite{JStat2006}, and the broad tail reflects the
heterogeneity in the resulting modular sizes. As a consequence, each
of these eigenvalues --with its associated large timescale, $t_i =
1/\lambda_i$-- contributes to the sum above, giving rise to a
convolution of relaxation processes, entailing anomalously-slow
dynamics, which could not be explained by a single-level modular
network (see Fig.3d-e): slow dynamics necessarily stems from the
existence of a hierarchy of moduli and structural bottlenecks.  As
explained in Methods, in the case of the HC the convolution of
different times scales gives rise  to stretched-exponential decay,
which perfectly fits with numerical results in Fig.3.
It was noted in the past that strongly modular networks exhibit
isolated eigenvalues in the lower edge of the laplacian
spectrum. Synchronization would develop in a step-wise process in
time, where each transient would be given by each isolated eigenvalue
\cite{Arenas_scales}.  In our case, the depth of the hierarchical
organization and the strength of topological disorder produce instead
a quasi-continuous tail of eigenvalues, and the step-wise process is
replaced by an anomalous stretched-exponential behavior.

\subsection*{A more refined model: hierarchical modular synthetic
  networks} To shed additional light on the previous findings for the
HC --i.e. the emergence of chimera-like states and anomalously slow
dynamics-- we suggest to go beyond the single-level modular network
model and study hierarchical modular networks (HMN) in which moduli
exists within moduli in a nested way at various scales
\cite{Bullmore-Sporns,Sporns,Review-Kaiser,Review-Bullmore,Zhou06,Ivkovic}.
HMN are assembled in a bottom-up fashion: local fully-connected moduli
(e.g. of $16$ nodes) are used as building blocks.  They are
recursively grouped by establishing additional inter-moduli links in a
level-dependent way as sketched in Fig.4(top) \cite{Nat-Comm,Zhou11}.
\begin{figure}[htbp]
  \centering
\includegraphics[width=12cm,angle=0]{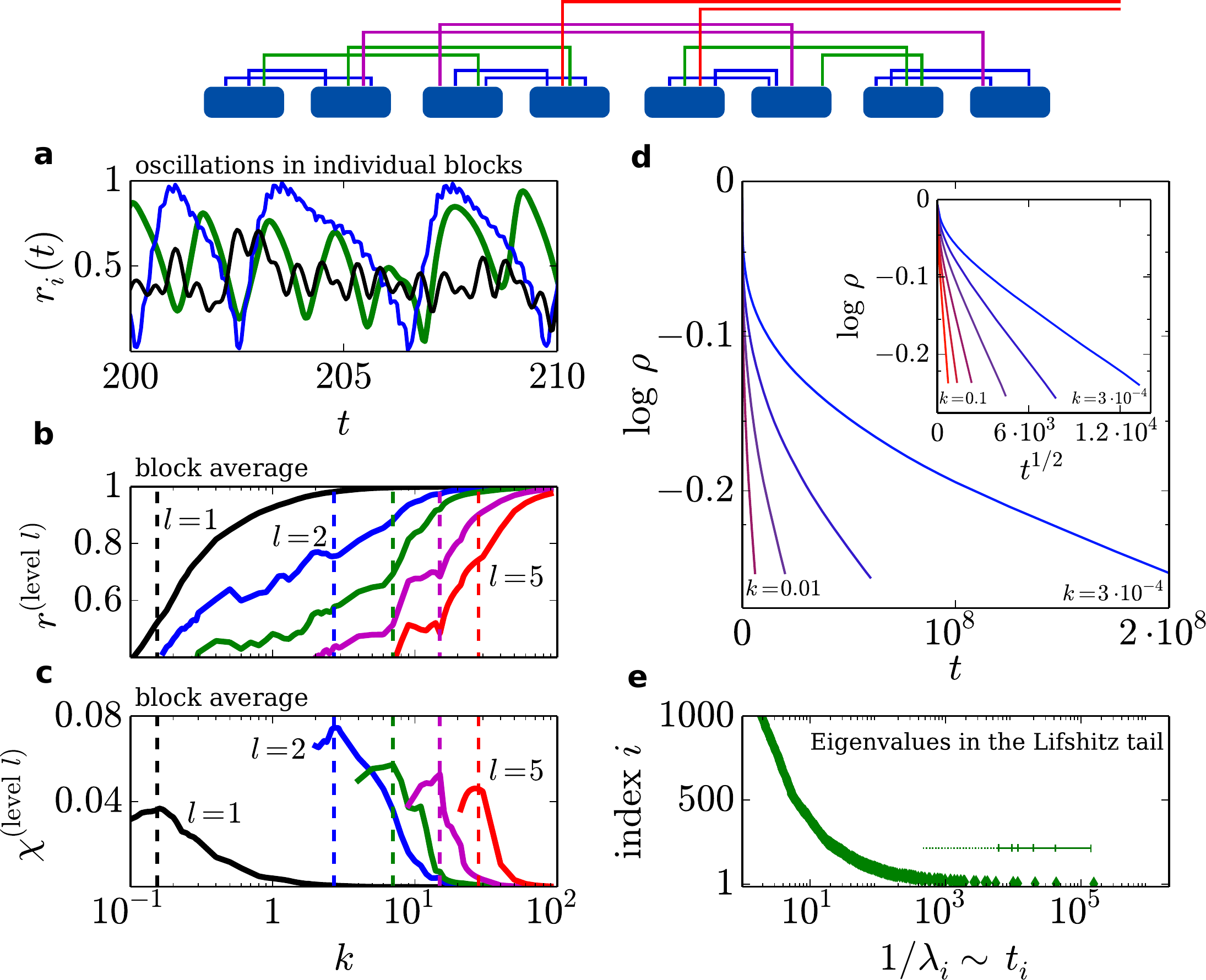}
\caption{Synchronization in hierarchical modular networks. 
  Top panel: sketch of the HMN
  model. At hierarchical level $1$, $2^s$ basal fully connected blocks
  of size $M$ are linked pairwise into super-blocks by establishing a
  fixed number $\alpha$ of random unweighted links between the
  elements of each ($\alpha=2$ in the Fig.). Newly formed blocks are
  then linked iteratively with the same $\alpha$ up to level $s$,
  until the network becomes connected.  a), b), c) as in Fig.3, but
  for a HMN with $N=512$, $s=5$, and $\alpha=4$.  Hierarchical levels
  are $i=1\to 5$ in black, blue, green, magenta and red respectively
  (not all shown in a) for clarity).  d) Time relaxation of activity
  $\rho$ for homogeneous characteristic frequencies $\omega=0$, for
  logarithmically equally spaced values of $k$.  Averages over $10^6$
  realizations of HMNs with $N=4096$ and $s=11$.  Inset: as in the
  main plot d), but representing as a function of $t^{1/2}$ and
  confirming the predicted stretched exponential behavior. e) Inverse
  tail-eigenvalues (as in Fig.3) for a HMN as in e).  
}
\end{figure}
Our computational analyses of the Kuramoto dynamics on HMN substrates
(see Fig.4) reveal: (i) a sequence of synchronization transitions for
progressively higher hierarchical levels at increasing values of $k$,
(ii) chimera-like states at every hierarchical level, resulting in a
hierarchy of metastable states with maximal variability at the
corresponding transition points, (iii) extremely slow relaxation
toward the coherent state when all internal frequencies are
identical. Furthermore, anomalies in the Laplacian spectrum analogous
to those of the HC network are observed for HMN matrices; in
particular, the lower edge of the HMN Laplacian spectrum has been
recently shown to exhibit a continuous exponential Lifshitz tail
$p(\lambda)\sim \mathrm{e}^{-1/\lambda^a}$ for $N \rightarrow \infty$,
with $a\approx 1$ \cite{Nat-Comm}.  Taking the continuum limit of
Eq.(\ref{eq:discrete_lambda}), we find $\rho(t)\approx
\frac{\sigma^2}{2}\int d\lambda\, p(\lambda) {e}^{-2 k\lambda t}$,
which can be evaluated with the saddle-point method (see Methods),
leading to
\begin{equation}\label{eq:stretched}
\rho(t) \sim \mathrm{e}^{- \sqrt{8 k t}},
\end{equation}
i.e. anomalous stretched-exponential asymptotic behavior, in excellent
agreement with computational results (see Fig.4d).  Therefore,
hierarchical modular networks constitute a parsimonious and adequate
model for reproducing all the complex synchronization phenomenology of
the HC. 

A crucial role in the emergence of such behavior is played by disorder.
One would be tempted to believe that all networks characterized by a finite
spectral dimension could potentially give rise to this phenomenology. This is 
obviously not the case for a regular lattice, where the spectral gap is always 
well defined. A fractal lattice or an ordered tree, on the other hand, could 
exhibit a hierarchy of discrete low eigenvalues, whose multiplicities reflect 
system symmetries. The introduction of disorder, as in HMNs, is then necessary 
in order to transform such hierarchy of discrete levels into a continuous 
Lifshitz tail, leading eventually to the behavior predicted by Eq. 
(\ref{eq:stretched}).

\section*{Discussion}

Simple models of synchronization dynamics exhibit an unexpectedly rich
phenomenology when operating on top of empirical human brain networks.
This complexity includes oscillatory behavior of the order parameter
suggesting the existence of relatively isolated structural communities
or moduli, that --as a matter of fact-- can be identified by using
standard community detection algorithms. Even more remarkably,
oscillations in the level of internal coherence are also present
within these moduli, suggesting the existence of a whole hierarchy of
nested levels of organization, as also found in the recent literature
relying on a variety of approaches
\cite{Bullmore-Sporns,Sporns,Review-Kaiser,Review-Bullmore,Buzsaki,Zhou06,
Ivkovic,Sporns2014}.
Aimed at unveiling this complex behavior we have introduced a family
of hierarchical modular networks and studied them in order to assess
what structural properties are required in order to reproduce the complex
synchronization patterns observed in brain networks.

In the absence of frequency dispersion, perfect coherence is achieved
in synthetic hierarchical networks by following a bottom-up ordering
dynamics in which progressively larger communities --with inherently
different timescales-- become coherent (see
\cite{Arenas_scales}). However, this hierarchically nested
synchronization process is constrained and altered by structural
bottlenecks --as carefully described here for the simpler two-block
toy model-- at all hierarchical levels.  This structural complexity
brings about anomalously-slow dynamics at very large timescales.
Observe that the HC, in spite of being a coarse-grained mapping of a
brain network, already shows strong signals of this ideal hierarchical
architecture as reflected in its anomalously slow synchronization
dynamics as well as in the presence of non-degenerate eigenvalues in
the lower edge of its Laplacian spectrum, acting as a fingerprint of
structural heterogeneity and complexity.  We stress that such a
complex phenomenology would be impossible to obtain in networks with
stronger connectivity patterns (e.g. with the small world property)
such as scale free-networks or high-degree random graphs. Even the
generic presence of simple communities may not be sufficient to grant
the emergence of frustration: the uniqueness of the human connectome,
and of hierarchical modular networks in general, resides in the strong
separation into distinct levels, which the synchronization dynamics is
able to resolve only at well-separated values of the coupling $k$.

On the other hand, in the presence of intrinsic frequency
heterogeneity, the described slow ordering process is further
frustrated. Actually, for small values of the coupling constant $k$
the system remains trapped into metastable and chimera-like states with
traits of local coherence at different hierarchical levels. In this
case, inter-moduli frequency barriers need to be overcome before
weakly connected moduli achieve mutual coherence. This is clearly
exemplified by the separation between distinct peaks in the chimera
index $\chi^{(l)}$ in Figs. 3-4, each one signaling the onset of an
independent synchronization process at a given level (see
Methods). The result is a complex synchronization landscape, which is
especially rich and diverse in the intermediate regime put forward
here.

Including other realistic ingredients such as explicit phase
frustration \cite{Shanahan2010} or time delays
\cite{Sporns2011,Shanahan2012} to our simplistic approach should only
add complexity to the structural frustration effect reported here. It
is also expected that more refined models --including neuro-realistic
ingredients leading to collective oscillations-- would generate
similar results, but this remains to be explored in future works.

Addition of noise to the Kuramoto dynamics would allow the system to
escape from metastable states. Stochasticity can overcome the
``potential barriers'' between mutually incoherent moduli as well as
re-introduce de-synchronization effects.  These combined effects can
make the system able to explore the nested hierarchy of attractors,
allowing one to shed some light into the complex synchronization
patterns in real brain networks. Actually, spontaneous dynamical
fluctuations have been measured in the resting state of human brains
\cite{RS}; these are correlated across diverse segregated moduli and
characterized by very slow fluctuations, of typical frequency $<0.1Hz$,
in close agreement with those found here (Fig.3).  Accordingly, it has
been suggested that the brain is routinely exploring different states
or attractors \cite{Deco2012} and that --in order to enhance
spontaneous switching between attractors-- brain networks should
operate close to a critical point, allowing for large intrinsic
fluctuations which on their turn entail attractor ``surfing'' and give
access to highly varied functional configurations
\cite{Chialvo10,Shew09,Deco2012,Haimovici,Shriki2013} and, in
particular, to maximal variability of phase synchrony \cite{Yang2012}.
  
The existence of multiple attractors and noise-induced surfing is
largely facilitated in the broad intermediate regime first elucidated
here, implying that a precise fine tuning to a critical point
might not be required to guarantee functional avantages usually
associated with criticality \cite{Chialvo10,Beggs08,Plenz2013}: the
role usually played by a critical point is assumed by a broad
intermediate region in hierarchically architectured complex systems
\cite{Nat-Comm}.  Finally, let us remark that our results might also
be of relevance for other hierarchically organized systems such as
gene regulatory networks \cite{Escherichia} for which coherent
activations play a pivotal role \cite{Kauffman08}.

\section*{Methods}
\subsection*{Numerical simulation of the Kuramoto model} 
The Kuramoto model is simulated by numerically integrating
Eq.~(\ref{eq:Kuramoto}). Computations are carried out using both a 4th
order Runge-Kutta method of fixed step size $h=10^{-3}$ and an 8th
order Dormand-Prince method with adaptive step size. Both methods lead
to compatible results within precision limits. The robustness of the
observation of a novel intermediate phase --between incoherent and
coherent ones-- is assessed by choosing different functional forms for
the frequency distribution $g(\omega)$ (Lorentzian, Gaussian,
uniform), and by implementing variations of Eq.~\ref{eq:Kuramoto} in
which the matrix ${\bf W}$ is weight-normalized \cite{Arenas-Review}
for simulations of the HC and degree-normalized for simulations in
HMNs. No qualitative change in the phenomenology is observed.

\subsection*{Chimera index $\chi^{(l)}$ and hierarchical synchronization}
In the main text, the chimera index $\chi^{(l)}$ is introduced as a measure of 
partial
synchronization at the community level $l$. At any hierarchical level $l$, a 
hierarchical network
can be divided into a set of communities. Following \cite{Shanahan2010}, 
$\chi^{(l)}$ is
defined as follows: (i) in the steady (oscillatory) state, and for each time 
$t$, local order 
parameters $r_i^{(l)}(t)$ for each community $i$ are calculated and their 
variance across
communities $\sigma_{\mathrm{chi}}^{(l)}(t)$
is stored; (ii) the chimera index is computed as the time average 
$\chi^{(l)}=\langle \sigma_\mathrm{chi}^{(l)}(t) \rangle_t$. Having 
$\chi^{(l)}>0$ at a given 
hierarchical level $l$  implies that local 
order is only partial as $r_i^{(l)}$ fluctuates, giving rise to a chimera-like 
state. On the other hand,
$\chi^{(l)}=0$  means that each local order parameter at that level is 
$r_i^{(l)}\approx 1$,
and local order has been attained. Figs. 3b-c and 4b-c show that at each $l$ 
(each color) a peak 
in the corresponding $\chi^{(l)}$ marks the onset of the local synchronization 
processes:
as soon as the peak vanishes upon increasing $k$, local order at that level is 
attained. 
The sequence of separated peaks in $\chi^{(l)}$ for increasing values of $l$ is 
the direct evidence 
of a hierarchical synchronization process.

\subsection*{Lifshitz tail and stretched-exponential asymptotic behavior}
In sparse HMNs, the lower end of the Laplacian spectrum is
characterized by an exponential tail in the density of states
$p(\lambda)\sim \mathrm{e}^{-1/\lambda^a}$, known as Lifshitz tail
\cite{Nat-Comm}.  In graphs, Lifshitz tails signal the existence of
non-trivial heterogeneous localized states governing the asymptotic
synchronization dynamics at very large times $t$. In the main text we
have shown that in the absence of frequency heterogeneity, the
$t\to\infty$ behavior of the activity is given by $\rho(t)\approx
\frac{\sigma^2}{2}\int d\lambda\, p(\lambda) {e}^{-2 k \lambda t}$.
This expression can be evaluated by applying the saddle point method,
yielding
\begin{equation}
\rho(t)\approx\frac{\sigma^2}{2}
\exp\left[ 
-(1+a)a^{-\frac{a}{1+a}}(2k t)^\frac{a}{1+a}
\right].
\end{equation}
Substituting $a\approx 1$, as empirically found in HMNs \cite{Nat-Comm}, 
leads to Eq.~(\ref{eq:stretched}),
whose square root behavior is confirmed by simulations in Fig. 4d.

\section*{Acknowledgements}
We acknowledge financial support from J. de Andaluc{\'\i}a, grant
  P09-FQM-4682 and we thank O. Sporns for providing us access to the
  human connectome data.

\section*{Author Contributions}
P.M. and M.A.M. conceived the project,
P.V. and P.M. performed the numerical simulations, 
carried out the analytical calculations and 
prepared the figures. 
P.M. and M.A.M. wrote the main manuscript text. 
All authors reviewed the manuscript.

\section*{Additional Information}
The authors declare no competing financial interests.


\begin{thebibliography}{10}

\bibitem{Hagmann}
Hagmann, P. {\it et al.} 
Mapping the structural core of human cerebral cortex. 
{\it PLoS Biol.} {\bf 6}, e159 (2008).

\bibitem{Honey09}
Honey, C. J. {\it et al.} 
Predicting human resting-state functional connectivity from
  structural connectivity.
{\it Proc. Natl. Acad. Sci. USA} {\bf 106}, 2035--2040 (2009).

\bibitem{Bullmore-Sporns}
Bullmore, E. \& Sporns, O. 
Complex brain networks: graph theoretical
  analysis of structural and functional systems.
{\it Nat. Rev. Neurosci.}
{\bf 10}, 186--198 (2009).

\bibitem{Sporns}
Sporns, O. 
{\it Networks of the Brain}.
(MIT Press, Cambridge, 2010).

\bibitem{Review-Kaiser}
Kaiser, M. 
A tutorial in connectome analysis: topological and spatial
  features of brain networks 
{\it NeuroImage} {\bf 57}, 892--907 (2011).

\bibitem{Review-Bullmore}
Meunier, D., Lambiotte, R. \& and Bullmore, E. 
Modular and hierarchically modular organization of brain networks.
{\it Front. Neurosci.} {\bf 4}, 200 (2010).

\bibitem{Buzsaki}
Buzs\'aki, G. 
{\it Rhythms of the Brain}.
(Oxford University Press, New York, 2006).

\bibitem{Zhou06}
Zhou, C., Zemanov\'a, L., Zamora, G.,  Hilgetag, C.~C. \& Kurths, J. 
Hierarchical organization unveiled by functional connectivity in 
  complex brain networks.
{\it Phys. Rev. Lett.} {\bf 97}, 238103 (2006).

\bibitem{Ivkovic}
Ivkovi{\'c}, M., Amy, K. \& Ashish, R. 
Statistics of weighted brain networks reveal hierarchical organization 
  and gaussian degree distribution.
{\it PLoS ONE} {\bf 7}, e35029 (2012).

\bibitem{Sporns2014}
Betzel, R. F. {\it et al.} 
Multi-scale community organization of the human structural connectome 
  and its relationship with resting-state functional connectivity.
{\it Network Science} {\bf 1}, 353--373 (2013).

\bibitem{Zhou07}
Zhou, C, Zemanov\'a L,  Zamora-L\'opez,  G., Hilgetag, C.~C. \& Kurths, J.
Structure--function relationship in complex brain networks expressed by
  hierarchical synchronization.
{\it New J. Phys.} {\bf 9}, 178 (2007).

\bibitem{Kaiser07}
Kaiser, M., G\"orner, M. \& Hilgetag, C.~C.
Criticality of spreading dynamics in hierarchical cluster networks without 
inhibition.
{\it New J. Phys.} {\bf 9}, 110 (2007).

\bibitem{Kaiser10}
Kaiser, M. \& Hilgetag, C~C. 
Optimal hierarchical modular topologies for producing
  limited sustained activation of neural networks.
{\it Front. Neuroinform.} {\bf 4}, 8 (2010).

\bibitem{Rubinov}
Rubinov, M., Sporns, O., Thivierge, J.~P. \& Breakspear, M. 
Neurobiologically realistic determinants of self-organized criticality in 
  networks of spiking neurons.
{\it PLoS Comput. Biol.} {\bf 7}, e1002038 (2011).

\bibitem{Nat-Comm}
Moretti, P. \& Mu\~noz, M.~A. 
Griffiths phases and the stretching of criticality in brain networks.
{\it Nat. Commun.} {\bf 4}, 2521 (2013).

\bibitem{Vojta-Review}
Vojta, T. 
Rare region effects at classical, quantum and nonequilibrium phase
  transitions.
{\it J. Phys. A} {\bf 39}, R143--R205 (2006).

\bibitem{GPCN}
Mu\~noz, M.~A., Juh\'asz, R., Castellano, C. \& \'Odor, G. 
Griffiths Phases on Complex Networks.
{\it Phys. Rev. Lett.} {\bf 105}, 128701 (2010).

\bibitem{GPCN2}
Juh\'asz, R., \'Odor, G., Castellano, C.,  \& Mu\~noz, M.~A. 
Rare-region effects in the contact process on networks.
{\it Phys. Rev. E} {\bf 85}, 066125, (2012).

\bibitem{Bennett2004}
Bennett, M. V. \& Zukin, R. 
Electrical coupling and neuronal synchronization in the mammalian brain.
{\it Neuron} {\bf 41}, 495--511 (2004).

\bibitem{Breakspear-multiscale}
Breakspear, M. \& Stam, C.~J. 
Dynamics of a neural system with a multiscale architecture.
{\it Phil. Trans. R. Soc. Lond. B} {\bf 360}, 1051--1074 (2005).

\bibitem{Sompolinsky88}
Sompolinsky, H., Crisanti, A. \& and Sommers, H. J. 
Chaos in random neural networks.
{\it Phys. Rev. Lett.} {\bf 61}, 259--262 (1988).

\bibitem{Klimesch1996}
Klimesch, W. 
Memory processes, brain oscillations and EEG synchronization.
{\it Int. J. Psychophysiol.} {\bf 24}, 61--100 (1996).

\bibitem{Deco-inf-transf}
Buehlmann, A. \& Deco, G. 
Optimal information transfer in the cortex through synchronization.
{\it PLoS Comput. Biol.} {\bf 6}, e1000934 (2010).


\bibitem{Niebur2000}
Steinmetz, P.~N. {\it et al.}
Attention modulates synchronized neuronal firing in primate 
  somatosensory cortex. 
{\it Nature} {\bf 404}, 187--190 (2000).

\bibitem{Kandel00}
Kandel, E.~R., Schwartz, J.~H. \&  Jessell, T.~M.
{\it Principles of Neural Science}.
(McGraw-Hill, New York, 2000)

\bibitem{RPK-book}
Rosenblum, M.~G., Pikovsky, A. \& Kurths, J.
{\it Synchronization -- A universal concept in nonlinear sciences}.
(Cambridge University Press, Cambridge, 2001).

\bibitem{Kuramoto75}
Kuramoto, Y. 
Self-entrainment of a population of coupled nonlinear oscillators.
{\it Lect. Notes Phys.} {\bf 39}, 420--422 (1975).

\bibitem{Strogatz00}
Strogatz, S.~H. 
From Kuramoto to Crawford: exploring the onset of
  synchronization in populations of coupled oscillators.
{\it Physica D} {\bf143}, 1--20 (2000).

\bibitem{Acebron-Review}
Acebr\'on, J.~A., Bonilla, L.~L., P\'erez Vicente, C.~J., Ritort, F., \&
  Spigler, R.
The Kuramoto model: a simple paradigm for synchronization phenomena.
{\it Rev. Mod. Phys.} {\bf 77}, 137--185 (2005).

\bibitem{Arenas-Review}
Arenas, A.,  D\'iaz-Guilera, A.,  Kurths, J., Moreno, Y. \& Zhou, C.
Synchronization in complex networks.
{\it Phys. Rep.} {\bf 469}, 93--153 (2008).

\bibitem{Sporns2011}
Cabral, J., Hugues, E., Sporns, O., \& Deco, G. 
Role of local network oscillations in resting-state functional connectivity.
{\it NeuroImage} {\bf 57}, 130--139 (2011).

\bibitem{Breakspear2010}
Breakspear, M., Heitmann, S. \& Daffertshofer, A.
Generative models of cortical oscillations: neurobiological implications 
of the Kuramoto model.
{\it Front. Hum. Neurosci.} {\bf 4}, 190 (2010).

\bibitem{Arenas10cat}
G{\'o}mez-Garde{\~n}es, J., Zamora-L{\'o}pez, G., Moreno, Y. \& Arenas, A.
From modular to centralized organization of synchronization in functional
  areas of the cat cerebral cortex,
{\it PLoS One} {\bf 5}, e12313 (2010).

\bibitem{OA}
Ott, E. \& Antonsen, T.~M. 
Low dimensional behavior of large systems of globally coupled oscillators.
{\it Chaos} {\bf 18}, 037113 (2008).

\bibitem{Skardal-Restrepo}
Skardal, P.~S. \& Restrepo, J. G. 
Hierarchical synchrony of phase oscillators in modular networks.
{\it Phys. Rev. E} {\bf 85}, 016208 (2012).

\bibitem{Arenas1}
Arenas, A. \& P\'erez-Vicente, C.~J.
Exact long-time behavior of a network of phase oscillators under random fields.
{\it Phys. Rev. E} {\bf 50}, 949--956 (1994).

\bibitem{Arenas2}
Acebr{\'o}n, J.~A. \& Bonilla, L.~L.
Asymptotic description of transients and
  synchronized states of globally coupled oscillators.
{\it Physica D} {\bf 114}, 296--314 (1998).

\bibitem{Popovych}
Popovych, O.~V., Maistrenko, Y.~L. \& Tass, P.~A.
Phase chaos in coupled oscillators.
{\it Phys. Rev. E} {\bf 71}, 065201 (2005).

\bibitem{Radatools}
Duch, J. \& Arenas, A.
Community detection in complex networks using extremal optimization.
{\it Phys. Rev. E} {\bf 72}, 027104 (2005).

\bibitem{Newman-Review}
Newman, M. 
The Structure and Function of Complex Networks. 
{\it SIAM Rev.} {\bf 45}, 167--256 (2003).

\bibitem{Abrams2008}
Abrams, D,~M.\& Strogatz, S.~H.
Chimera States for Coupled Oscillators.
{\it Phys. Rev. Lett.} {\bf 93}, 174102 (2004).

\bibitem{Arenas_scales}
Arenas, A., D\'iaz-Guilera, A. \& P\'erez-Vicente, C.~J.
Synchronization reveals topological scales in complex networks.
{\it Phys. Rev. Lett.} {\bf 96}, 114102 (2006).

\bibitem{Abrams2004}
Abrams, D,~M., Mirollo, R., Strogatz, S.~H. \& Wiley, D.~A.
Solvable model for chimera states of coupled oscillators.
{\it Phys. Rev. Lett.} {\bf 101}, 084103 (2008).

\bibitem{Shanahan2010}
Shanahan, M. 
Metastable chimera states in community-structured oscillator networks.
{\it Chaos} {\bf 20}, 013108 (2010).

\bibitem{Shanahan2012}
Wildie, M. \& Shanahan, M. 
Metastability and chimera states in modular delay and pulse-coupled oscillator 
networks.
{\it Chaos} {\bf 22}, 043131 (2012).

\bibitem{McGraw2005}
McGraw, P.~N. \& Menzinger, M.
Clustering and the synchronization of oscillator networks.
{\it Phys. Rev. E} {\bf 72}  015101(R) (2005).

\bibitem{FanChung}
Chung, F.~R.~K.  
{\it Spectral graph theory}.
(Reg. Conf. Series. in Maths, AMS, Providence, 1997).

\bibitem{JStat2006}
Donetti, L.,  Neri, R. \& Mu\~noz, M.~A. 
Optimal network topologies: expanders, cages, Ramanujan graphs, 
  entangled networks and all that.
{\it J. Stat. Mech.}, P08007 (2006).

\bibitem{Zhou11}
Wang, S.-J., Hilgetag, C.~C. \& Zhou, C. 
Sustained activity in hierarchical modular neural networks: SOC and 
oscillations.
{\it Front. Comput. Neurosci.} {\bf 5}, 30 (2011).

\bibitem{RS}
Biswal, B., Zerrin~Yetkin, F., Haughton, V. \& Hyde, J.  
Functional connectivity in the motor cortex of resting human brain using 
echo-planar
  MRI.
{\it Magnet. Reson. Med.} {\bf 34}, 537--541 (1995).

\bibitem{Deco2012}
Deco, G. \& and Jirsa, V.~K. 
Ongoing cortical activity at rest: criticality, multistability, and ghost 
attractors.
{\it J. Neurosci.} {\bf 32}, 3366--3375 (2012).

\bibitem{Chialvo10}
Chialvo, D.~R. 
Emergent complex neural dynamics.
{\it Nat. Phys.} {\bf 6} 744--750 (2010).

\bibitem{Shew09}
Shew, W.~L., Yang, H., Petermann, T., Roy, R. \& Plenz, D. 
Neuronal avalanches imply maximum dynamic range in cortical networks at 
criticality.
{\it J. Neurosci.} {\bf 29}, 15595--15600 (2009).

\bibitem{Haimovici}
Haimovici, A., Tagliazucchi, E.,  Balenzuela, P. \& Chialvo, D.~R.
Brain organization into resting state networks emerges at 
  criticality on a model of the human connectome.
{\it Phys. Rev. Lett.}  {\bf 110}, 178101 (2013).

\bibitem{Shriki2013}
Shriki, O. {\it et al.}
Neuronal avalanches in the resting MEG of the human brain.
{\it J. Neurosci.} {\bf 33}, 7079--7090 (2013).

\bibitem{Yang2012}
Yang, H., Shew, W.~L., Roy, R. \& Plenz, D.
Maximal variability of phase synchrony in cortical networks with neuronal 
avalanches
{\it J. Neurosci.} {\bf 32}, 1061--1072 (2012).

\bibitem{Beggs08}
Beggs, J.~M. 
The criticality hypothesis: how local cortical networks might
  optimize information processing.
{\it Phil. Trans. R. Soc. A} {\bf 366}, 329--343 (2008).

\bibitem{Plenz2013}
Shew, W.~L. \& Plenz, D. 
The functional benefits of criticality in the cortex.
{\it Neuroscientist} {\bf 19}, 88--100 (2013).

\bibitem{Escherichia}
{Trevi\~no III}, S., Sun, Y., Cooper, T.~F. \& Bassler, K. 
Robust detection of hierarchical communities from Escherichia coli gene 
expression data.
{\it PLoS Comput. Biol.}  {\bf 8}, e1002391 (2012).

\bibitem{Kauffman08}
Nykter, M. {\it et al.} 
Gene expression dynamics in the macrophage exhibit criticality.
{\it Proc. Natl. Acad. Sci. USA} {\bf 105}, 1897--1900 (2008).

\end{thebibliography}
\end{document}